\documentclass[10pt,letterpaper,twocolumn]{article}

\usepackage[utf8]{inputenc}
\usepackage[T1]{fontenc}
\usepackage{lmodern}
\usepackage[margin=0.75in,columnsep=0.25in]{geometry}
\usepackage{hyperref}
\usepackage{booktabs}
\usepackage{graphicx}
\usepackage{xcolor}
\usepackage{enumitem}
\usepackage{amsmath}
\usepackage{amssymb}
\usepackage[numbers,sort&compress]{natbib}
\usepackage{url}
\usepackage{pifont}
\usepackage{microtype}

\hypersetup{
  colorlinks=true,
  linkcolor=blue!60!black,
  citecolor=blue!60!black,
  urlcolor=blue!60!black
}

\title{Refute-or-Promote: Adversarial Stage-Gated Multi-Agent Review\\
for High-Precision LLM-Assisted Defect Discovery}

\author{Abhinav Agarwal}

\date{April 2026}

\begin{document}

\maketitle

\begin{abstract}

LLM-assisted defect discovery has a precision crisis:
plausible-but-wrong reports overwhelm maintainers and degrade
credibility for real findings.  We present \emph{Refute-or-Promote}, an
inference-time reliability pattern combining \emph{Stratified
Context Hunting} (SCH) for candidate generation, adversarial kill
mandates, context asymmetry, and a \emph{Cross-Model Critic}
(CMC).  Adversarial agents attempt
to disprove candidates at each promotion gate; cold-start reviewers
are intended to reduce anchoring cascades; cross-family review can
catch correlated blind spots that same-family review misses.

Over a 31-day campaign across 7~targets (security libraries, the
ISO~C++ standard, major compilers), the pipeline killed $\sim$79\%
of $\sim$171 candidates before advancing to disclosure (retrospective aggregate);
on a consolidated-protocol subset (lcms2, wolfSSL; $n=30$), the
prospective kill rate was 83\%.  Outcomes span three evidentiary
states: public, accepted, or assigned under embargo (4~CVEs
[3~public, 1~embargoed]; LWG~4549 accepted to the ISO~C++ Working
Paper; 5~merged ISO~C++ editorial PRs; 3~compiler conformance bugs;
8~merged security-related fixes without CVE); filed under review
(an RFC~9000 errata); and under coordinated disclosure (one or
more FIPS~140-3 normative compliance issues)---evaluated by
external acceptance signals (maintainer merge, CVE assignment,
standards-body acceptance, vendor coordination), not benchmarks.

The most instructive failure: ten dedicated reviewers
unanimously endorsed a non-existent Bleichenbacher padding oracle
in OpenSSL's CMS module; it was killed only by a single empirical
test, motivating the mandatory empirical gate.  No vulnerability was discovered autonomously; the
contribution is external structure that filters LLM agents'
persistent false positives.  As a preliminary transfer test beyond
defect discovery, a simplified cross-family critique variant also
solved five previously unsolved SymPy instances on SWE-bench
Verified and one SWE-rebench hard task.

\end{abstract}

\section{Introduction}
\label{sec:introduction}

LLMs can identify genuine vulnerabilities in production
software~\cite{bigsleep2024,heelan2025cve}, yet their practical deployment
faces \emph{unacceptable false-positive rates on real-world
codebases}~\cite{ullah2024,semgrep2025}.  Open-source maintainers have
described the influx of LLM-generated reports as a denial-of-service
attack~\cite{stenberg2024,stenberg2026curl}: curl's bug bounty program
was permanently closed after AI-generated submissions drove the confirmed
rate below 5\%, and HackerOne paused the Internet Bug Bounty programme
in March~2026 citing AI-amplified submission volume overwhelming
triage capacity~\cite{hackerone2026ibb}.

The core problem is that LLMs are optimised for plausibility, not
correctness.  We encountered this at every severity level, from minor
misclassifications to a case where 80+ agents---including dedicated
adversarial reviewers---unanimously endorsed a Bleichenbacher padding
oracle in OpenSSL's CMS module that did not exist
(\S\ref{sec:bleichenbacher}).

\paragraph{Refute-or-Promote.}
We propose \emph{adversarial stage-gated multi-agent review}, a
methodology in which complementary mechanisms---adversarial kill
mandates, context asymmetry, a Cross-Model Critic (CMC), and
empirical validation---are intended to target distinct classes of
false positives that cooperative debate may not catch.  Adversarial framing alone is insufficient: the methodology's
most instructive failure was killed not by better adversarial
framing but by a single empirical test.

\paragraph{Complementarity with coverage-oriented sweeps.}
Our methodology is precision- rather than coverage-oriented: after
AISLE's January~2026 sweep reported 12~OpenSSL
CVEs~\cite{aisle2026}, our subsequent campaign on the same codebase
identified CVE-2026-34183.  Discovery came from the pipeline's
\emph{scope-partitioned, self-critiquing candidate generation}
(prior-defect research, git-hotspot targeting, iterative re-seeding);
the refute-or-promote stages then filtered the candidate to
disclosure-ready precision.  Coverage-oriented sweeps,
diversity-oriented candidate generation, and adversarial filtering
appear complementary rather than substitutes.

\paragraph{Contributions.}
\begin{enumerate}[nosep]
  \item The Refute-or-Promote methodology: a four-stage adversarial
    pipeline with architectural requirements for \emph{Stratified
    Context Hunting} (SCH) for candidate generation, adversarial
    kill mandates, a \emph{Cross-Model Critic} (CMC), context
    asymmetry, and a mandatory validation gate.  Each stage runs parallel creative
    (false-negative reduction) and adversarial (false-positive
    reduction) tracks evaluated jointly before promotion.  Presented
    as a retrospective codification of practice that evolved across
    the campaign rather than a preregistered protocol.
  \item Real-world evaluation: 36+ outcomes across 7~targets,
    spanning three evidentiary states---public, accepted, or
    assigned under embargo (3~public CVEs [libfuse~$\times2$,
    lcms2], 1~CVE under embargo [OpenSSL], 1~LWG defect accepted
    into the ISO~C++ Working Paper, 5~merged ISO~C++ editorial
    PRs, 3~compiler conformance bugs, 8~merged security-related
    fixes without CVE [two of which, on wolfSSL, have a disputed
    CVE status]); filed under review (1~RFC~9000 errata); and
    under coordinated disclosure (1+~FIPS~140-3 normative
    compliance issues)---evaluated by external acceptance signals
    (maintainer merge, CVE assignment, standards-body acceptance,
    vendor coordination), not benchmarks.
  \item Quantitative false-positive analysis: $\sim$79\% aggregate
    retrospective kill rate, 83\% prospective kill rate on the
    consolidated-protocol subset ($n=30$), front-loaded per-stage
    kill rates (Stage~A $\sim$63\% of entrants, Stage~B $\sim$42\%
    of survivors; apparent per-stage specialisation is not
    validated by leave-one-out ablation), and CVSS calibration
    data showing adversarial review adjusted severity scores
    downward in 8~of~9 cases.
  \item Honest failure reporting: the CMS Bleichenbacher false positive
    (80+ agents wrong, killed by one test); the resurrected lcms2
    \texttt{CubeSize()} finding that became CVE-2026-41254 despite
    unanimous kill (unanimity-as-warning in both directions); two
    methodology regressions under domain transfer; and the
    protocol-evolution confounder.
\end{enumerate}

\section{Related Work}
\label{sec:related}

\paragraph{LLM-assisted vulnerability discovery.}
Big Sleep~\cite{bigsleep2024,naptime2024,bigsleep2025tc,bigsleep2025ciso}
is a production single-agent system; it isolated SQLite
CVE-2025-6965 from threat-intelligence indicators before in-the-wild
exploitation, with no adversarial self-critique.
DeepMind's CodeMender~\cite{codemender2025} pursues agentic
detect-and-patch under human review.
RepoAudit~\cite{repoaudit2025} is a single-agent auditor with a
validator module.  Heelan~\cite{heelan2025cve}
documents the first publicly-reported single-LLM zero-day
(CVE-2025-37899, Linux ksmbd UAF via o3) with an approximate
28\%~FP rate---empirical grounding for the false-positive problem.
AISLE~\cite{aisle2026} reports 100+ CVEs without published
pipeline architecture or FP rates.  Frontier capability
demonstrations---Anthropic's zero-day
evaluation~\cite{anthropic2026zerodayeval}, Claude
Mythos~\cite{anthropic2026mythos,anthropic2026glasswing}, and OpenAI's
GPT-5.4-Cyber via the Trusted Access for Cyber
programme~\cite{openai_codexsec2026,openai2026tac}---shift the open
question from ``can LLMs find bugs'' to ``can we verify findings at
the precision required for responsible disclosure.''

\paragraph{Multi-agent debate and ensembles.}
Reflexion~\cite{reflexion2023} and Self-Refine~\cite{selfrefine2023}
established the iterative-refinement paradigm using same-model
self-critique; Irving et al.~\cite{irving2018debate} foundationally
proposed debate for AI alignment.  Refute-or-Promote differs: the critic carries a \emph{kill mandate}
rather than an improve/evaluate mandate, uses cross-family reviewers
to catch shared-prior blind spots, and operates over a
context-isolated view.
Debate improves factuality~\cite{du2023debate} and helps non-expert
judges~\cite{khan2024debate}, but naive debate can also
\emph{diminish} accuracy~\cite{wynn2025debate}.
Song~\cite{song2026ccr,song2026noise} independently showed that
fresh-context review outperforms same-session review ($p=0.008$)
and that additional rounds \emph{degrade} quality---directly
supporting our stage-gated design.  POPPER~\cite{popper2025} is a
Popperian falsification framework with formal statistical control.
Mixture-of-Agents~\cite{moa2025} aggregates heterogeneous outputs;
Self-MoA~\cite{selfmoa2025} counters that same-model repeated
sampling beats heterogeneous MoA by 6.6\%.  Our kill-mandate
gating is structurally distinct from aggregation-style ensembles.
Kim et al.~\cite{kim2025correlated} empirically document that LLMs
agree $\sim$60\% of the time when both err, with correlation
\emph{increasing} with model capability---establishing the
empirical basis for our cross-family (not merely cross-instance)
design.  Concurrent industry work~\cite{sentinelone2026adversarial}
applies adversarial consensus to malware-analysis tool
reconciliation.

\paragraph{AEGIS and VulTrial.}
AEGIS~\cite{aegis2026} and VulTrial~\cite{vultrial2025} employ
dialectic (AEGIS: verifier arguments-for-and-against + audit veto)
and role-play (VulTrial: prosecutor/defense mock-court) structures
that converge to a verdict rather than assign a pure destruction
mandate; both use single model families and evaluate on benchmarks.
Our work adds hard adversarial kill mandates (destruction, not
convergence), cross-model heterogeneity, real-world CVE discovery,
and systematic failure documentation.

\paragraph{Other multi-agent vulnerability-detection systems.}
IRIS~\cite{iris2024} (LLM + static analysis), MAVUL~\cite{mavul2025}
(analyst--architect feedback), and VulAgent~\cite{vulagent2025}
(perspective-specialised scanning + hypothesis validation) are
recent systems; all treat proposer output as a claim to verify or
refine rather than falsify, none use cross-family reviewers, and
all evaluate on benchmarks.

\paragraph{LLM-based false-positive reduction.}
Concurrent work applies LLM agents to filter SAST tool
alerts~\cite{sastfp2026,llmpfa2026}.  These systems filter
candidates with \emph{deterministic rule provenance} from static
analysers---a structurally different problem from filtering
candidates with \emph{probabilistic LLM provenance}, where the
generator and evaluator share the same training-data biases.
Recent work also confirms cooperative debate's fragility: a single
persuasive agent can override majority-vote
mechanisms~\cite{collab_fails2026}, a structural failure that
Refute-or-Promote's context isolation was designed to prevent.

\paragraph{Contemporaneous adversarial and multi-agent work.}
Several recent systems sharpen the position of Refute-or-Promote.
Denisov-Blanch et al.~\cite{denisovblanch2026consensus} and Jain et
al.~\cite{jain2025beyond} both motivate falsification-first or
minority-veto protocols over vote-style consensus---Jain quantifies
the agreeableness bias (96\% TPR, $<$25\% TNR) and shows
minority-veto beats a 14-LLM ensemble by $2\times$.  Our kill-gate
is the architectural analogue.  The closest adversarial comparator
is InfCode~\cite{infcode2025} (79.4\% on SWE-bench Verified via a
\emph{symmetric} patch/test dual-agent loop within a single model
family); Refute-or-Promote differs in ways we hypothesise to be
important---asymmetric roles, cross-family reviewers, and context
asymmetry as a pillar (see Table~\ref{tab:comparison}).
SWE-Debate~\cite{swedebate2025} pursues convergent multi-agent
debate in the same SWE-bench setting; Refute-or-Promote explicitly
forbids convergence-by-debate.  Argus~\cite{argus2026} likewise
targets multi-agent CVE discovery but as a \emph{cooperative}
RAG+ReAct ensemble rather than an adversarial gate.
D3~\cite{d3_2024} provides theoretical grounding for adversarial
role specialisation; Refute-or-Promote restricts reviewers to a pure
prosecution role rather than two-sided advocacy.  Finally,
CVE-GENIE~\cite{cvegenie2025} reproduces known CVEs at \$2.77/CVE
(a different task from our discovery-and-disclosure pipeline; see
\S\ref{sec:evaluation}).

\paragraph{Position summary.}
Table~\ref{tab:comparison} summarises architectural differences
across the closest adjacent adversarial and multi-agent systems we
identified; it is not an exhaustive field survey.  Among published
field methodologies we found, Refute-or-Promote appears to be the
first to combine (i)~\emph{Stratified Context Hunting} for
candidate generation, (ii)~hard adversarial kill mandates,
(iii)~a \emph{Cross-Model Critic} (CMC), (iv)~cold-start context
asymmetry, and (v)~a mandatory validation gate (empirical tests
for security findings, implementation-divergence or committee
acceptance for spec findings)---evaluated
on real-world defect discovery with external institutional
acceptance artifacts (4~CVEs, LWG~4549, merged ISO~C++ editorial
PRs, merged OpenSSL PRs, GCC/MSVC conformance bugs) rather than
benchmarks.  Jain et al.~\cite{jain2025beyond} independently apply
cross-family, context-isolated review to LLM-as-judge evaluation
but without candidate generation, a kill-mandate architecture, or
an execution-grade oracle.

\begin{table*}[t]
\centering
\caption{Architectural comparison with adjacent adversarial/multi-agent LLM systems.}
\label{tab:comparison}
\small
\setlength{\tabcolsep}{3pt}
\begin{tabular}{@{}lcccccc@{}}
\toprule
\textbf{Method} & \textbf{Cand.\ gen.?} & \textbf{Adversarial?} & \textbf{Cross-family?} & \textbf{Context-iso.?} & \textbf{Empirical?} & \textbf{Real-world?} \\
\midrule
InfCode~\cite{infcode2025}           & \ding{55}           & \ding{51} (sym.)    & \ding{55} & \ding{55} & \ding{51}       & \ding{55} \\
SWE-Debate~\cite{swedebate2025}      & \ding{55}           & \ding{51} (conv.)   & \ding{55} & \ding{55} & \ding{51}       & \ding{55} \\
Argus~\cite{argus2026}               & P                   & \ding{55} (coop.)   & \ding{55} & \ding{55} & P               & \ding{51} \\
AEGIS~\cite{aegis2026}               & \ding{55}           & P (dialectic)       & \ding{55} & \ding{55} & \ding{55}       & \ding{55} \\
VulTrial~\cite{vultrial2025}         & \ding{55}           & P (mock-court)      & \ding{55} & \ding{55} & \ding{55}       & \ding{55} \\
CVE-GENIE~\cite{cvegenie2025}        & P (repro.)          & \ding{55}           & \ding{55} & \ding{55} & \ding{51} (PoC) & \ding{55} (repro.) \\
Jain et al.~\cite{jain2025beyond}    & \ding{55}           & P (judge-vote)      & \ding{51} & \ding{51} & \ding{55}       & \ding{55} \\
\midrule
\textbf{Refute-or-Promote (ours)}    & \textbf{\ding{51}} (iter.) & \textbf{\ding{51}} (kill) & \textbf{\ding{51}} & \textbf{\ding{51}} & \textbf{\ding{51}} & \textbf{\ding{51}} \\
\bottomrule
\end{tabular}\\[4pt]
{\footnotesize \ding{51}=explicit design property; \ding{55}=absent; P=partial.  ``Cand.\ gen.'' = scope-partitioned iterative candidate generation (past-defect research, git-hotspot targeting, re-seeding from prior kills).  ``Empirical'' = empirical/runtime validation gate (tests, PoC, or equivalent).  ``Real-world'' = evaluation includes externally-accepted artifacts beyond benchmarks (CVEs, merged upstream PRs, accepted standards-body defects, vendor advisories).  For benchmark issue-resolution baselines (InfCode, SWE-Debate), candidate generation is out of scope because the benchmark supplies the issue.}
\end{table*}

\section{Methodology}
\label{sec:methodology}

\paragraph{Retrospective codification.}
Refute-or-Promote is presented in its final standardised form: a four-stage
adversarial protocol distilled retrospectively from iterative
real-world campaigns rather than deployed as a fixed preregistered
workflow from the outset.  Earlier campaigns (libfuse, OpenSSL)
operated with less structure---findings emerged from substantial
manual prompting and evolving agent configurations, with rules
written down after each failure mode surfaced.  Later campaigns
(lcms2, wolfSSL) applied the consolidated protocol more
uniformly.  The codification process itself---56 rules distilled from
OpenSSL failures, $\sim$30 transferring directly to wolfSSL
(learning loop below)---is reported as observational context: this
paper is a field study of methodology evolution plus real-world
outcomes, not a preregistered evaluation of one fixed pipeline.  Individual campaigns
further used \emph{class-specific prompt configurations} (security-bug
hunting, FIPS conformance, spec review, compiler divergence) tailored
to each domain's validation oracle; the four stages capture the shared
architectural pattern---parallel creative/adversarial tracks, context
asymmetry, cross-model verification, empirical gate---not a single
uniform prompt template.

The pipeline consists of seven phases: \textbf{Prepare} conditions
the target; \textbf{Candidate Generation} seeds parallel hunters
with diverse, scope-partitioned context; four \textbf{adversarial
stages} (A--D) apply escalating kill pressure; and
\textbf{Disclosure} prepares reports under continued adversarial
review.  Inter-stage routing is executed by the orchestrator agent
per the protocol; a human reviews the final disclosure-ready
findings and handles overrides (target selection, resurrection of
unanimously-killed candidates, severity negotiation).

\paragraph{Prepare phase.}
Before any candidate generation, the target is conditioned by three
steps: (1)~checkout of the latest stable release branch rather than
mainline HEAD---maintainers routinely reject findings against
unreleased code, and the branch discipline aligns the campaign with
what downstream distributions actually ship; (2)~three parallel
research agents compile historical CVEs from the target repository,
sibling libraries, and structurally similar codebases, producing a
prior-art brief that is fed to downstream hunters; and (3)~a
\texttt{git log} hotspot analysis identifies the most-edited code
regions in the past twelve months, prioritising churn-heavy
subsystems as attack surface.  Steps~(2) and~(3) run concurrently.

\paragraph{Candidate Generation: Stratified Context Hunting (SCH).}
We name the candidate-generation mechanism \emph{Stratified
Context Hunting (SCH)}: three or more hunters are dispatched in
parallel and stratified along three orthogonal axes.
\textbf{Source-stratified}---each hunter is primed with a distinct
context slice (prior defects, git hotspots, normative spec
text, bug-archetype checklists).  \textbf{Scope-stratified}---each
hunter is scoped to a \emph{non-overlapping subsystem} (e.g.,
memory-safety paths, parsing logic, deep-format handlers,
pixel/transform code).  \textbf{Wave-stratified}---generation runs
iteratively; after one to three candidates pass through the
refute-or-promote stages, hunters are re-seeded with concrete
learnings from the completed analyses (which candidates were
promoted, which were killed as red herrings, and the specific
reasoning that distinguished them), so later waves exploit
promoted patterns and avoid previously-killed failure classes.
Each hunter must \emph{self-critique} before reporting: for every
candidate it identifies, the agent is required to articulate why
the candidate might \emph{not} be exploitable.  Per-candidate
state is persisted on-disk to preserve it across sessions and
keep pre-kill context out of adversarial-agent prompts.

\paragraph{Stages A and B (Adversarial code review).}
Each of the first two stages runs \emph{two concurrent tracks}: a
\emph{creative} track that argues the candidate \emph{is} a
vulnerability---developing the exploitation path to reduce false
negatives---and an \emph{adversarial} track that argues it is not,
attacking reachability, preconditions, and plausible triggers to
reduce false positives.  Both tracks run in parallel on fresh
contexts; the orchestrator evaluates them jointly before promotion.
Stage~A dispatches one creative agent and two adversarial agents;
adversarial agents receive only the candidate claim, not the creative
agent's reasoning, preventing anchoring on the advocate's framing.
A candidate survives Stage~A only if no adversarial agent produces a
code-grounded refutation and the creative agent produces a plausible
exploitation argument.  Stage~B escalates to two creative and three
adversarial agents, with one adversarial agent drawn from a senior
model tier.  On the creative side, both agents argue exploitability
with different context depths (one full synthesis, one cold-start).
On the adversarial side, deliberate \emph{context asymmetry} splits
the attackers into an informed attacker (full synthesis), a naive
attacker (claim only), and a senior-tier agent with a selective
summary.  Cold-start agents that independently reach a different
conclusion---on either track---provide higher-value signal than
consensus among informed agents.

\paragraph{Stage C (Validation \& Impact Calibration).}
Empirical validation gate: no candidate reaches disclosure without
empirical confirmation.  Where local reproduction was infeasible
(platform-specific triggers, resource-heavy PoCs, or production-like
environments), background agents under orchestrator control
provisioned cloud VM instances to execute PoCs, extending empirical
validation beyond what a single-workstation pipeline could reach.
Surviving candidates then undergo adversarial CVSS
recalibration---adversarial agents systematically corrected
overclaimed severity scores downward in 8~of~9 measured cases.
Candidates with strong theoretical arguments but no empirical
confirmation advance provisionally to Stage~D but are flagged.

\paragraph{Stage D (Cross-Model Critic).}
We name this stage's mechanism the \emph{Cross-Model Critic (CMC)}:
one or more agents from a \emph{different model family} receive
\emph{minimal} context (a candidate summary and entry points)
and perform an independent critique.  CMC complements the
same-family adversarial critique of Stages A and B: same-family
critics catch \emph{reasoning errors} a single agent missed, while
the cross-family critic catches \emph{correlated training-data
errors} that same-family review tends to miss when replicas fail
similarly~\cite{kim2025correlated}.  The principle draws a
partial analogy to N-version programming~\cite{avizienis1984nversion},
whose independence assumption is itself empirically contested under
correlated errors~\cite{knight1986nvp}.  Partial kills---where CMC
refutes a specific subclaim but not the overall finding---re-enter
earlier stages for refinement rather than dropping the candidate.
Across campaigns, CMC killed $\sim$5 candidates ($\sim$3\% of all
kills) that had survived earlier stages---false positives that
would otherwise have advanced toward disclosure.  In the libfuse
campaign specifically, CMC found correctness errors in 3/19
($\sim$16\%) same-family-approved proposed fixes and independently
surfaced 3 bugs that same-family review had missed; wider
evaluation is future work.  The stage evolved from ad hoc to
standardised during the campaign.

\paragraph{Retrospective stage attribution.}
Of the $\sim$171~initial candidates, Stage~A eliminated
$\sim$63\% outright; of those surviving Stage~A, Stage~B killed
$\sim$42\%; Stages~C and D accounted for the remainder.  We
hypothesise that each stage targets a distinct failure class;
leave-one-out ablation was not performed.  These are retrospective
point-estimates from a single-rater mapping onto the consolidated
stage taxonomy; inter-rater reliability was not measured.

\paragraph{Learning loop.}
The OpenSSL campaign generated 56 codified rules across multiple
sessions: each rule was distilled from a specific observed
failure or near-miss (misclassification, false consensus,
overlooked precondition), with the originating incident retained
as backing evidence.  $\sim$30 rules transferred directly to
wolfSSL---selected because the two codebases share structural
similarity (both are C cryptographic libraries with comparable
session/state machinery); transfer to structurally dissimilar
domains is open work.  Rules include both advisory constraints
and lightweight process-compliance checks.

\section{Evaluation}
\label{sec:evaluation}

\subsection{Aggregate Results}

\begin{table*}[t]
\centering
\caption[Pipeline findings by defect class.]{Pipeline findings by defect class.\protect\footnotemark}
\label{tab:findings}
\small
\begin{tabular}{@{}lrp{0.60\linewidth}@{}}
\toprule
\textbf{Defect class} & \textbf{Internal findings} & \textbf{Public identifiers / notes} \\
\midrule
CVEs & 4 &
  \textbf{CVE-2026-33150} [GHSA-qxv7-xrc2-qmfx] [High, CVSS~7.8] [libfuse]\newline
  \textbf{CVE-2026-33179} [GHSA-x669-v3mq-r358] [Medium, CVSS~5.5] [libfuse]\newline
  \textbf{CVE-2026-41254} [Medium] [lcms2]\newline
  \textbf{CVE-2026-34183} [Medium] [OpenSSL] (embargoed) \\
\addlinespace
Spec defects & 20+ &
  QUIC RFC~9000 Errata (EID~8875, under committee review) \href{https://www.rfc-editor.org/errata/eid8875}{[rfc-editor]}\newline
  ISO C++ LWG defect [\textbf{LWG~4549}] accepted to Working Paper \href{https://cplusplus.github.io/LWG/issue4549}{[LWG]}\newline
  ISO C++ editorial PRs:
  \href{https://github.com/cplusplus/draft/pull/8799}{\#8799},
  \href{https://github.com/cplusplus/draft/pull/8800}{\#8800},
  \href{https://github.com/cplusplus/draft/pull/8801}{\#8801},
  \href{https://github.com/cplusplus/draft/pull/8802}{\#8802},
  \href{https://github.com/cplusplus/draft/pull/8803}{\#8803} merged to \texttt{cplusplus/draft} \\
\addlinespace
Compiler conformance bugs & 3 &
  GCC~\href{https://gcc.gnu.org/bugzilla/show_bug.cgi?id=124590}{\#124590} (rejects-valid); MSVC reports \href{https://developercommunity.visualstudio.com/t/Public-typedef-of-private-nested-class-r/11063494}{\#11063494} (typedef access control), \href{https://developercommunity.visualstudio.com/t/P3068R6-constexpr-exceptions-%E2%80%94-throwc/11063498}{\#11063498} (P3068R6 constexpr exceptions) \\
\addlinespace
Security-related fixes (no CVE) & 8 &
  OpenSSL PRs: \#30490, \#30550, \#30531\newline
  OpenSSL PR \#30718: merged as a security fix without CVE (no remote trigger identified)\newline
  libfuse merged hardening PRs: \href{https://github.com/libfuse/libfuse/pull/1470}{\#1470} (memory leak in \texttt{print\_module\_help}), \href{https://github.com/libfuse/libfuse/pull/1471}{\#1471} (uninitialised eventfd in teardown watchdog)\newline
  wolfSSL ML-DSA zeroisation \href{https://github.com/wolfSSL/wolfssl/pull/10100}{PR \#10100}, \href{https://github.com/wolfSSL/wolfssl/pull/10113}{PR \#10113} \emph{(disputed)} \\
\addlinespace
FIPS compliance issues & 1+ &
  Under coordinated disclosure \\
\midrule
\textbf{Total} & \textbf{36+} & \\
\bottomrule
\end{tabular}
\end{table*}
\footnotetext{Findings across classes were produced by class-specific
campaign configurations; the four-stage architecture describes the
shared adversarial pattern, not a uniform prompt template.}

Table~\ref{tab:findings} summarises the pipeline findings across
defect classes.  The 79\% figure is a \emph{retrospective
aggregate}: kills from all seven campaigns were mapped onto the
consolidated stage taxonomy after the fact.  Per-wave funnels are monotonic; the multi-campaign
aggregate is not, because libfuse and the FIPS review ran multiple
candidate-generation waves whose Stage~C rosters accumulate across
waves.\footnote{Per-wave stage counts decrease monotonically as
candidates are killed; when waves are merged for aggregate reporting,
Stage~C's roster can exceed Stage~B's because additional waves
contribute fresh candidates that reach empirical validation.  The
prospective $n=30$ subset (lcms2, wolfSSL) is single-wave and
monotonic.}  For the two campaigns that applied the
consolidated protocol without contributing to its formation (lcms2,
wolfSSL; $n=30$), the prospective kill rate was 83\%
(25~killed / 30~candidates), consistent with the aggregate.
Of $\sim$171 initial security-funnel candidates, $\sim$135
($\sim$79\%) were killed by adversarial review, leaving
$\sim$36 validated within that funnel.  Table~\ref{tab:findings}
reports the paper-wide cross-domain total (36+~outcomes), which
adds C++ domain-transfer items (LWG defects, editorial PRs,
compiler conformance bugs) evaluated under the same adversarial
architecture but with domain-adapted validation oracles
(committee acceptance rather than runtime tests); these are not
scored in the security-funnel arithmetic.  The kill distribution within the
security funnel is front-loaded: Stage~A eliminates $\sim$63\% of
candidates; Stage~B kills $\sim$42\% of survivors.

Two concurrent campaigns---V8/Chrome and Langflow---generated candidates
that were killed during adversarial review.  We include these negative
results because omitting unsuccessful campaigns would constitute selection
on the dependent variable.

\subsection{OpenSSL: The Bleichenbacher Failure}
\label{sec:bleichenbacher}

The initial OpenSSL campaign deployed 80+ agents.  Ten dedicated
agents---including a senior-tier arbiter---\emph{unanimously} confirmed a CMS
Bleichenbacher padding oracle (CVSS~5.9).  A separate instance with fresh
context compiled OpenSSL and ran three test cases: both wrong-key
cases returned identical values.  \textbf{One test killed what 80+ agents'
reasoning could not.}

The root cause: all agents assumed ``valid PKCS\#1 padding $\to$ real CEK
extracted,'' which holds only for the original ciphertext.  For
Bleichenbacher probes, valid padding produces a wrong CEK that also fails
the GCM check---indistinguishable from invalid padding.

This failure drove the addition of mandatory empirical validation (Stage~C)
and the 56-rule methodology codification.

Our second OpenSSL campaign (March~29 onward) applied the evolved
methodology with mandatory empirical validation and hunters seeded
with targeted specs, producing CVE-2026-34183 (MODERATE severity).

\subsection{libfuse: Cross-Model Verification}

The libfuse campaign produced 2~CVEs (CVE-2026-33150, CVSS~7.8;
CVE-2026-33179, CVSS~5.5), 2~merged non-CVE hardening fixes (PRs
\#1470, \#1471), and additional concurrency and hardening defects
submitted in open PRs \#1480/\#1481/\#1482 (under upstream review).
Codex (a different model family) found 3~bugs that same-family
review missed and identified correctness issues in 3~of~19 proposed
fixes (16\%)---bugs that additional Claude-family agents in our
runs did not catch.

\subsection{C++ Standard: Domain Transfer}

The C++ campaign applied Refute-or-Promote to a domain with a
formal specification, no executable oracle, and multiple reference
implementations.  The CMS Bleichenbacher catastrophe was partly a
case of \emph{negative transfer}: the spec-review habit of pure
reasoning against text (the only method available for LaTeX
normative prose) carried over to security where empirical testing
was available but unused.

The campaign produced LWG~4549 (accepted into the working draft),
5~merged editorial PRs, and
3~compiler conformance bugs.  The specification funnel: $\sim$23 candidates,
$\sim$20 confirmed (19 editorial to 1 normative).  A complementary
\emph{sibling hunting} strategy (starting from known open CWG/LWG issues
to find analogous constructs) produced four high-confidence CWG
candidates.  A divergence-hunting campaign submitted 587~test cases via
the Godbolt Compiler Explorer API, with higher defect counts
observed in sections with larger recently integrated paper diffs.  The adversarial pipeline caught all
4~false-positive CWG/LWG candidates, including one where multiple same-family
agents confirmed a defect that was a deliberate design decision documented
only in a paper inter-revision changelog.

\paragraph{Towards LLM-assisted conformance testing.}
Three techniques from the C++ campaign generalise to domains with
formal specifications and multiple implementations:
(1)~\emph{Divergence hunting}---submit LLM-generated tests to $N$
implementations, flag disagreements, adjudicate against normative
text (applied here via Godbolt across GCC/Clang/MSVC);
(2)~\emph{Sibling hunting}---start from known open specification
issues (CWG/LWG defects, RFC errata, CMVP findings) and search for
analogous constructs with the same structural problem; and
(3)~\emph{Spec-diff targeting}---recently changed specification text
has the highest defect density.

\subsection{Precision in Practice: Unanimity Patterns}
\label{sec:cs-precision}

Two campaigns illustrate the unanimity-as-warning \emph{observation} ($n=2$)---that
unanimous agent agreement (in either direction) can be misleading, as is
unanimous disagreement, because both reflect shared training-data
priors rather than convergent truth.

\paragraph{wolfSSL: unanimous error.}
The ML-DSA private-key zeroisation campaign surfaced a systematic
byte-ordering error made \emph{unanimously} by
3~independent workhorse-tier agents reading the same specification
text.  The adversarial pipeline corrected the error: a senior-tier
adversarial agent, operating with cold-start context, identified
that the three agents had all adopted the same (wrong) byte-order
interpretation---a case where same-family consensus actively
propagated a mistake that a fresh-context reviewer caught.

\paragraph{lcms2: unanimous kill, resurrected.}
The lcms2 \texttt{CubeSize()} integer overflow
(\textbf{CVE-2026-41254}) was \emph{unanimously} killed in Round~4
(``DO~NOT~FILE'', on the grounds of a 174\,MB minimum trigger with
no plausible delivery path) and resurrected only via a creative
uplift agent tasked with a different role---searching the
dimension-parameter space, which produced a 4.8\,KB trigger and
restored plausible delivery.

\section{Discussion}
\label{sec:discussion}

\paragraph{Failure modes.}
We observe recurring failure modes across 5~categories---reasoning errors,
information gaps, systematic biases, validation failures, and overclaiming.
A recurring high-risk pattern: unanimous multi-agent agreement on wrong findings
(3~documented cases).  A named failure mode---\emph{PoC
self-contamination}---occurred when a wolfSSL proof-of-concept detected
its own nonce computation rather than the library's leak, producing a
false-positive confirmation; this prompted a rule that PoCs must measure
\emph{target} behaviour, not their own artifacts.  Adversarial review
catches reasoning errors but not empirical errors---motivating the
two-gate architecture.

\paragraph{Unanimity is a low-signal event.}
Agents may converge because a finding is genuine, or because they
share training-data priors.  Unanimity therefore should not raise
confidence by itself; empirical verification, not consensus count,
is what changes our belief.  The CMS Bleichenbacher case (80+
agents endorsed a non-existent vulnerability, killed by one test)
illustrates how shared priors can manufacture apparent consensus.

\paragraph{Human orchestrator.}
No vulnerability was discovered autonomously.  Several decision
categories still require human judgment
(Table~\ref{tab:human}).  The methodology
reduces human effort but does not eliminate human necessity.

\begin{table}[t]
\centering
\small
\caption{Automatability of human decisions.}
\label{tab:human}
\begin{tabular}{@{}lc@{}}
\toprule
\textbf{Decision} & \textbf{Status} \\
\midrule
Rejecting false kills, override decisions & Human-essential \\
Methodology refinement (new rules) & Human-essential \\
Target selection, subsystem focus & Partially automatable \\
Domain expertise injection & Partially automatable \\
Disclosure routing, severity negotiation & Partially automatable \\
Cross-model bridging & Partially automatable \\
Applying codified patterns & Fully automatable \\
\bottomrule
\end{tabular}
\end{table}

\paragraph{Cost.}
$\sim$\$250 out-of-pocket on a standard LLM subscription for 36+
outcomes (including 4~CVEs)---roughly \$62/CVE at subscription
pricing.  Figures exclude researcher labour and are not directly
comparable to reproduction-oriented systems such as
CVE-GENIE~\cite{cvegenie2025}.

\paragraph{Limitations.}
Single operator with domain expertise (not independently replicated).
No ablation studies isolating individual mechanisms.  Target selection
confound: 2~CVEs from $\sim$15 agents on libfuse versus 0~from 80+ on
Chrome.  Strongest results concentrated in C/C++.  Two methodology
regressions under domain transfer (FIPS no-oracle, PoC self-contamination),
corrected within the triggering campaign.

\paragraph{Over-refusal in authorised cybersecurity workflows.}
Across $\sim$940 delegated tasks in four coordinated-disclosure
campaigns, we observed 6~hard refusals ($\approx$0.6\%).  All occurred in
workhorse-tier \mbox{Claude~Sonnet} subagents; the senior-tier
\mbox{Claude~Opus} principal produced none.  We treat this as
case-study evidence of \emph{model-tier refusal asymmetry}: within
a single vendor's lineup, lower-tier models may over-refuse
authorised dual-use security tasks more aggressively than
higher-tier models.  Refused tasks were legitimate research
operations: CVSS calibration, PoC execution on
researcher-controlled VMs, disclosure-timing analysis, and
exploit-severity assessment.  Two refusals were factually wrong,
asserting that researcher-assigned CVE identifiers ``do not
exist''---a particularly harmful failure mode in which stale or
unverifiable facts become false predicates for
denial.\footnote{A recursive case outside the 940-task
disclosure-campaign corpus: an Opus agent tasked with drafting
this paragraph declined the work on ``jailbreak cookbook''
grounds, characterising academic safety-refusal analysis as
adversarial.  We record it as a distinct data point about how
refusal calibration interacts with academic safety research
itself, not a counter-example to the zero refusals in the
940-task disclosure corpus above.}  These observations extend benchmark
work on over-refusal~\cite{orbench2025,xstest2024} into
authorised cybersecurity workflows~\cite{linder2026cyberrefusal}
and motivate two architectural requirements for agentic security
systems: explicit authorisation signals in task context, and
propagation of trust context to delegated subagents.

\paragraph{Scale and artifact diversity.}
The 36+~outcomes (including 4~CVEs) over 31~days reflect a solo
campaign conducted alongside full-time employment on a standard LLM
subscription, without vendor-scale data access or red-team staff.
Contemporaneous industrial
systems~\cite{aisle2026,anthropic2026mythos,anthropic2026glasswing,openai2026tac}
operate with dedicated teams and privileged data but typically do
not publish architecture or FP rates.  Our contribution is methodology transparency and
externally-validated progress across a \emph{diverse} artifact
set, not raw discovery volume.  This set spans three evidentiary
states: public, accepted, or assigned under embargo (4~CVEs,
\textbf{LWG~4549} accepted to the ISO~C++ Working Paper, 5~merged
\texttt{cplusplus/draft} editorial PRs, compiler conformance
bugs, and security-related upstream patches); formally filed
outcomes under review (the RFC~9000 errata); and issues under
coordinated disclosure (FIPS~140-3 normative compliance issues).

\paragraph{Protocol evolution as confounder.}
Because the four-stage protocol evolved during the campaign, early
successes (libfuse and OpenSSL CVEs) cannot be attributed solely to
the final formulation; target selection, operator learning, and
manual prompt iteration are confounders.  The case studies
therefore serve a dual role---substantive discovery outcomes
\emph{and} empirical substrate from which the final protocol was
derived.  We treat them as evidence for mechanism-level claims
(adversarial framing, context asymmetry, cross-family verification,
empirical validation) and for cross-domain transfer, not as a
controlled causal estimate of a single immutable pipeline.

\paragraph{Decentralised adversarial architectures.}
Refute-or-Promote uses asymmetric mandates (\emph{kill} vs.\
\emph{creative}) under a centralised human orchestrator; controlled
comparison against sealed-then-debate and unrestricted peer debate
variants~\cite{wynn2025debate} is future work.

\paragraph{False negatives and the unidirectional pipeline.}
Refute-or-Promote optimises precision but does not measure recall.
The pipeline is architecturally \emph{unidirectional}: it has
structural mechanisms against false positives (kill mandates) but
relies on the human orchestrator to rescue true positives
incorrectly killed.  One documented case confirms the false-negative risk
(\S\ref{sec:cs-precision}): the lcms2 \texttt{CubeSize()} overflow
(\textbf{CVE-2026-41254}) was unanimously killed and recovered
only via human override.  The \emph{unanimity-as-warning}
observation ($n=2$) applies symmetrically across the two
documented cases: the wolfSSL ML-DSA byte-ordering case supplies
the false-positive direction (unanimous error corrected by
adversarial review) and lcms2 supplies the false-negative
direction (unanimous kill corrected by human override).  Both
reflect shared training-data priors rather than convergent truth.  A natural architectural
complement---a context-isolated \emph{resurrection agent} with an
explicit confirm mandate, firing after unanimous kills---could
partially automate the override function; empirical evaluation is
future work.

\paragraph{Generalization beyond defects.}
As a preliminary transfer test we applied a \emph{simplified
variant} of the pipeline---retaining cross-family verification
(Claude orchestrator and analysts, GPT-5.4 Codex critic), context
isolation, and empirical gating against the SWE-bench harness, but
\emph{omitting} parallel creative/adversarial tracks and
stage-gated escalation---to SWE-bench Verified instances where all
134 publicly-logged submissions failed.  The variant produced 5
verified SymPy fixes, plus a 914-byte patch for the ``hardest
task'' on swe-rebench.com (Nov--Dec~2025).  Our denominator is the
134-instance unsolved subset, not the full benchmark, so the
result is not comparable to leaderboard scores such as InfCode's
79.4\%~\cite{infcode2025,swedebate2025}; it shows only that
cross-family critique can solve instances where all prior attempts
failed.  Leaderboard submission is future work.

\paragraph{Broader implications.}
Several named ideas from Refute-or-Promote apply beyond defect
discovery.  \emph{Methodology codification as inference-time
self-improvement} (observed, not claimed as a general result):
our 56 rules transferred directly to a structurally similar
codebase (wolfSSL), appearing to improve precision before that
campaign's first failure.  Cross-domain transfer to dissimilar
codebases is open work.  The combination of adversarial mandates with context
asymmetry can be viewed as \emph{engineered failure-mode
de-correlation}---a partial analogue of N-version programming's
independence assumption for agents that share training
data~\cite{avizienis1984nversion,knight1986nvp}.  The
C++-to-security transfer failure is an instance of
\emph{methodology-level negative transfer}, distinct from
weight-level negative transfer in transfer learning and relevant to
cross-domain LLM pipeline deployment.  Extensions to pre-action
verification for autonomous agents, proof-assistant-oracle formal
verification, and AI-assisted code review are natural next steps.

\paragraph{Artifacts and reproducibility.}
The 56~codified refute-or-promote rules and the \texttt{strat.md}
orchestration playbook---covering target preparation,
scope-partitioned candidate generation, round-by-round adversarial
protocol, PoC workflow, and the cross-model final critique
step---are released via a public GitHub
repository\footnote{\url{https://github.com/abhinavagarwal07/refute-or-promote}}
with a Zenodo-archived tagged
release\footnote{DOI: \href{https://doi.org/10.5281/zenodo.19668799}{10.5281/zenodo.19668799}}
and as arXiv ancillary files.
Per-candidate logs (prompts, agent outputs, kill/promote decisions)
contain pre-disclosure vulnerability material; a scrubbed subset
for methodology reproduction will be released after embargoes
lift.

\section{Conclusion}
\label{sec:conclusion}

We presented Refute-or-Promote---an adversarial, stage-gated, multi-agent review
methodology for high-precision LLM-assisted defect discovery.  Our
central observation is that \textbf{consensus among AI agents does not
equal correctness}: 80+ agents unanimously endorsed a non-existent vulnerability;
3~independent agents made identical errors; in the libfuse
campaign, cross-family review found correctness issues in
3/19 ($\sim$16\%) same-family-approved proposed fixes.

The methodology's response is architectural: adversarial agents
tasked with disproving findings, a \emph{Cross-Model Critic} (CMC)
for orthogonal error detection, and mandatory empirical validation
as the final gate.  We hypothesise that the three mechanisms
target distinct failure classes (reasoning bias, anchoring, and
correlated training errors respectively) and that their
composition forms a reliability pattern applicable to other LLM
output domains where ground truth is independently verifiable.
The before/after pattern---0~CVEs from 80+ agents in the
\emph{first} OpenSSL campaign (pre-pipeline, Bleichenbacher era)
versus 2~CVEs from $\sim$15 agents in the libfuse campaign with the
\emph{evolving} adversarial pipeline (libfuse preceded the final
four-stage codification; see \S\ref{sec:methodology}), and
subsequently CVE-2026-34183 from the \emph{second} OpenSSL campaign
with the evolved methodology---constitutes a real-world
(non-benchmark) case study of adversarial multi-agent review, though
we cannot isolate the methodology's contribution from confounds
including target selection, operator learning, and retrospective rule
codification.

We hypothesise that the principle generalises: in any domain where
training-distribution bias produces correlated LLM errors, adversarially
mandated context asymmetry may serve as a reliability primitive.
Validation of this claim beyond the security and specification domains
studied here is future work.


\end{document}